\documentclass[aps,prl,superscriptaddress,twocolumn]{revtex4}
\usepackage{graphicx}
\usepackage{amsmath}

\newcommand{\be}{\begin{equation}}
\newcommand{\ee}{\end{equation}}
\newcommand{\bea}{\begin{eqnarray}}
\newcommand{\eea}{\end{eqnarray}}

\newcommand{\unit}[1]{\ensuremath{\;{\rm #1}}}
\newcommand{\kmps}{\unit{km/s}}
\newcommand{\keV}{\unit{keV}}
\newcommand{\GeV}{\unit{GeV}}

\newcommand{\erf}{\ensuremath{\operatorname{erf}}}

\begin{document}

\title{Assessing Alternatives for Directional Detection of a WIMP Halo}

\author{Craig J. Copi}
\email{cjc5@cwru.edu}
\affiliation{Center for Education and Research in Cosmology and Astrophysics, Department of Physics}
\author{Lawrence M. Krauss}
\email{lmk9@cwru.edu}
\affiliation{Center for Education and Research in Cosmology and Astrophysics, Department of Physics}
\affiliation{Department of Astronomy \\
Case Western Reserve University,
10900 Euclid Ave., Cleveland OH 44106-7079}
\author{David Simmons-Duffin}
\affiliation{Physics Dept, Harvard University}
\author{Steven R. Stroiney}
%%\email{srs20@po.cwru.edu}
\affiliation{Physics Department, Cornell University}

\date{\today}

\begin{abstract}
  The future of direct terrestrial WIMP detection lies on two fronts: new, much larger low background detectors sensitive to energy deposition, and detectors with directional sensitivity.  The former can probe a large range of WIMP parameter space using well tested technology while the latter may be necessary if one is to disentangle particle physics parameters from astrophysical halo parameters.  Because directional detectors will be quite difficult to construct it is worthwhile exploring in advance generally  which experimental features will yield the greatest benefits at the lowest costs.  We examine the sensitivity of
  directional detectors with varying angular tracking resolution with and without the ability to distinguish forward versus backward recoils, and compare these to the sensitivity of a
  detector where the track is projected onto a two-dimensional plane.   The latter detector regardless of
  where it is placed on the Earth, can be oriented to produce a
  significantly better discrimination signal than a 3D
  detector without this capability, and with sensitivity within a factor of 2 of a full 3D tracking detector.   Required event rates to distinguish signals from backgrounds for a simple isothermal halo range from the low teens in the best case to many thousands in the worst.
\end{abstract}

%insert suggested PACS numbers in braces on next line
%%\pacs{95.35.+d}
\pacs{}

% Crude hack to get preprint number to showup in non preprint mode
%\makeatletter
%\@booleantrue\preprint@sw
%\makeatother

%\preprint{CWRU-PXX-XX}

\maketitle

\section{Introduction}

Direct detection experiments for weakly interacting massive particles
(WIMPs) continue to set ever more stringent limits on the nucleon-WIMP cross
section~\cite{dama-results,cdms1-results,edelweiss-results,zeplin-results,cdms2-first}.
A
precise understanding of the backgrounds is required to identify an excess
of nuclear recoils.  There is no unique signature that can separate neutron
induced recoils from WIMP induced recoils in these detectors.  Even annual modulation, which is at best a few percent effect, might be accounted for by seasonal background variations.  (This is for a pure WIMP signal.  When a uniform
background is included the effect is even smaller.)  It has been recognized
that a stronger signal comes from measuring the direction of the recoiling
nucleus~\cite{spergel} allowing for a WIMP signal to be identified from a
few events even in the presence of backgrounds~\cite{chk,ck}.
Detectors that might measure the recoil direction
have been designed and built.  For example, time projection chambers have
been used by DRIFT~\cite{drift} and NEWAGE~\cite{newage} and a
scintillator with direction dependent response has also been
studied~\cite{scintillator}.

In their present state the directional detectors are rudamentary at best.
 The current version of DRIFT, for example,
does not measure the full three dimensional track of the recoiling nucleus.
Instead it measures the recoil track projected onto a plane and has a
number of other limitations.  A detailed statistical study of DRIFT's
current capabilities has been performed~\cite{directional-statistics}  

A new generation of directionally sensitive detectors are now being
envisaged.  Building a full three dimensional detector is a challenging,
costly proposition.  Is it necessary?  
Given the technical challenge and cost constraints of these detectors,
which changes will lead to the most sensitive detector?  Here we provide a
general analysis of various design goals to determine the number of events
required for detector designs ranging from two dimensional to fully three
dimensional detectors.  We do not focus on any particular detector
technology.  Instead we apply a consistent set of parameters to a variety
of detector configurations.  This allows us to determine the optimal design goals.  In
the process we also are able to dispel various myths about optimal locations on 
Earth for detector placement, and the relative efficacy of two-dimensional versus three-dimensional event reconstruction.

\section{Theoretical Model}

\subsection{Detector Characteristics}

To quantify the capabilities of directionally sensitive detectors we
consider a consistent, generic set of parameters.  For the detector target
we use a xenon ($m_N=131\GeV$) nucleus.  We assume a threshold of
$Q_{\textrm{th}} = 10\keV$ and two different WIMP masses, $m_\chi=100\GeV$
($m_\chi \approx m_N$) and $m_\chi=1000\GeV$ ($m_\chi\gg m_N$).  For our
Galaxy we focus solely on an isothermal model for the WIMP halo
distribution,
\begin{equation}
  f (\vec v) = \frac1{\pi^{3/2}v_0^3} e^{-\left.|\vec v|^2\right/v_0^2}.
\end{equation}
We study three different values for $v_0$ spanning the range of current
expectations, $170\kmps$, $220\kmps$, and $270\kmps$.  The escape velocity
of WIMPs from the Galaxy is taken to be $650\kmps$.  The Earth's rotation
axis is oriented at an angle $\epsilon=42^\circ$ with respect to the Sun's
motion.  This value is relevant for the two dimensional detector  We stress
that these choices have been made to provide a consistent set of parameters
to allow the intercomparison of detector designs.  Thus, what will be important about our results will not be absolute constraints, but relative ones, although we expect the overall order of magnitude for the required event rates will not differ compared to our estimates.

\subsection{Differential Event Rates}

The technique for calculating the WIMP scattering rate is well known~\cite{chk}.  
The differential rate as a
function of nuclear recoil direction $(\gamma, \phi)$ is given by
\begin{equation}
  \frac{dR}{d\Omega_{\gamma,\phi}} = \frac{\sigma_0\rho_0}{\pi m_N
    m_\chi} \int_{\cal R} d^3\vec v\, J v\, F^2\! \left(Q (Jv)\right) f (\vec
  v+\vec v_\oplus).
  \label{eqn:3Drate}
\end{equation}
Here $m_N$ is the mass of the target nucleus, $m_\chi$ is the WIMP mass,
$\vec v_\oplus$ is the velocity of the Earth through the WIMP halo,
$Q$ is the recoil energy of the nucleus,
$\sigma_0$ is the cross section for WIMP scattering off the target nucleus,
and $\rho_0$ is the local WIMP halo density.  We consider only spin
independent interactions and use the standard Helm form factor~\cite{helm}
for $F^2 (Q)$.  The geometry of the WIMP scattering gives
\begin{equation}
 Jv \equiv v_x \sin\gamma\cos\phi + v_y\sin\gamma\sin\phi + v_z\cos\gamma.
\end{equation}
The integration region, $\cal R$, is defined by the detector threshold,
$Q_{\rm th}$, at the lower limit,
\begin{equation}
  Jv \ge \sqrt{\frac{\left (m_\chi+m_{N}\right)^2}{2m_\chi^2 m_N} Q_{\rm
      th}},
\end{equation}
and the galactic escape velocity, $v_{\rm esc}$, at the upper limit,
\begin{equation}
  \left(v_x+v_{\oplus,x}\right)^2 + \left(v_y+v_{\oplus,y}\right)^2 +
  \left(v_z+v_{\oplus,z}\right)^2 \le v_{\rm esc}^2.
\end{equation}
See~\cite{chk} for a more detailed discussion.  A full three dimensional
detector probes the reaction rate outlined here~(\ref{eqn:3Drate}).

Measuring the direction the nucleus is moving along a track is not always
possible.  For a three dimensional detector without forward-backward
discrimination, a recoil in a direction $(\cos\gamma,\phi)$ cannot be
distinguished from a recoil in a direction $(-\cos\gamma,\phi+\pi)$.  The
event rate for these two directions thus combine giving a total rate for
the direction $(\cos\gamma,\phi)$ of $\frac{dR}{d\Omega} (\cos\gamma,\phi)
+ \frac{dR}{d\Omega} (-\cos\gamma,\phi+\pi)$ where the differential rates
are again given above~(\ref{eqn:3Drate}).

A two-dimensional detector can only resolve the recoil direction projected
onto a plane.  We assume for comparison purposes however that it can be designed with forward-backward discrimination.  Suppose the normal to the detector plane is oriented at angles
$(\zeta,\eta)$ with respect to the direction of the Sun's motion.  The rate
is a function of a single angle $\phi'$ measured in this frame,
\begin{equation}
  \frac{dR}{d\phi'} = \frac{\sigma_0\rho_0}{\pi m_N m_\chi} \int_{-1}^1
  d\left (\cos\gamma'\right) \int_{\cal R} d^3\vec v' (Jv)' F^2 (Q') f
  (\vec v+\vec v_\oplus) .
\end{equation}
Here, primed coordinates are measured in the frame of the detector (where
the normal vector points along the $z$-axis) and unprimed coordinates are
measured in the frame of the Sun's motion.  These two frames are related by
rotations through the angles $\zeta$ and $\eta$.

For a detector fixed to the Earth's surface, the detector orientation with
respect to the Sun's motion changes throughout the day due to the Earth's
rotation.  Let $\theta$ be the angle between the detector's normal and the
Earth's rotation axis and $\epsilon$ be the angle between the rotation axis
and the Sun's velocity.  Then
\begin{equation}
  \cos\zeta = \cos\epsilon \cos\theta + \sin\epsilon\sin\theta \cos (2\pi
  t)
\end{equation}
for $t$ measured in days.

It is also important to recognize that even full three-dimensional detectors will not have perfect angular resolution.  To model
realistic angular resolution either due to dispersion of the recoiling
nucleus along its ideal track or due to inherent precision of the inherent detection mechanism itself we convolve the
ideal scattering rate~(\ref{eqn:3Drate}) with a smoothing kernel
$K(\Omega,\Omega')$,
\begin{equation}
  \frac{dR}{d\Omega}(\Omega) = \int K(\Omega,\Omega')
  \frac{dR}{d\Omega}(\Omega') d\Omega'.
\end{equation}
We use a Gaussian smoothing about the direction of the ideal recoil,
\begin{equation}
  K(\Omega,\Omega') =
  e^{-\xi^2/2\sigma^2}/(2\pi)^{3/2}\sigma\erf(\sqrt{2}\sigma),
  \label{eqn:smoothing-kernel}
\end{equation}
where
\begin{equation}
  \cos\xi = \sin\theta \sin\theta'\cos(\phi-\phi') + \cos\theta
  \cos\theta'.
\end{equation}
We study this as a function of the width of the Gaussian.

\section{Results}

We test the capabilities of each type of WIMP detector by assessing
their ability to distinguish the WIMP distribution from a flat background, using an isothermal halo as a fiducial test model.
The probability that a WIMP will recoil in a particular direction, $\Omega_i$,
is given by $P (\Omega_i) = \frac{dR}{d\Omega} (\Omega_i)$, where we have
normalized the rate such that $R=1$.  Thus, we are probing the shape of the
recoil spectrum.  The likelihood function for $N_e$ detected events is
defined by $\displaystyle {\cal L} = \prod_{i=1}^{N_e} P (\Omega_i)$.  We
generate at least 100,000 sample
distributions for each $N_e$ and apply the log-likelihood test to find
the minimum number of WIMP events such that we have a 95\%
detection 95\% of the time (see~\cite{chk} for more details).

\begin{table}
  \caption{The number of events required to identify a WIMP signal above a
    flat background for different types of detectors and a WIMP mass of
    $m_\chi=100\GeV$.}
  \label{tab:Nevents-100}
  \begin{ruledtabular}
    \begin{tabular}{lccc}
      \multicolumn{1}{c}{Detector} &
      \multicolumn{3}{c}{$v_0\;\mathrm{(km/s)}$} \\ \cline{2-4}
      \multicolumn{1}{c}{Type} & $170$ & $220$ & $270$ \\ \hline
      3D (full) & $6$ & $11$ & $18$ \\
      3D without FB & $176$ & $1795$ & $>35,000$ \\
      2D---best/worst & $19/45$ & $34/75$ & $61/123$ \\
      2D rotating & $13$ & $24$ & $43 $
    \end{tabular}
  \end{ruledtabular}
\end{table}

\begin{table}
  \caption{The number of events required to identify a WIMP signal above a
    flat background for different types of detectors and a WIMP mass of
    $m_\chi=1000\GeV$.}
  \label{tab:Nevents-1000}
  \begin{ruledtabular}
    \begin{tabular}{lccc}
      \multicolumn{1}{c}{Detector} &
      \multicolumn{3}{c}{$v_0\;\mathrm{(km/s)}$} \\ \cline{2-4}
      \multicolumn{1}{c}{Type} & $170$ & $220$ & $270$ \\ \hline
      3D (full) & $14$ & $27$ & $51$ \\
      3D without FB & $152$ & $217$ & $371$ \\
      2D fixed---best/worst & $51/129$ & $97/217$ & $175/368 $ \\
      2D rotating & $31$ & $61$ & $125 $
    \end{tabular}
  \end{ruledtabular}
\end{table}

The results for
$m_\chi=100\GeV$ are given in Table~\ref{tab:Nevents-100} for the range of
detectors we have considered, where `full' reflects a full three-dimensional detector with perfect angular resolution.  We shall discuss the degradation implied by limited resolution shortly.  The same set of results for
$m_\chi=1000\GeV$ are given in Table~\ref{tab:Nevents-1000}.  
Although we
have restricted our quantitative study to isothermal models the qualitative features of the comparison remain valid for other models, including models dominated by single streams of WIMPs .

Our results underscore the need for forward-backward detection.  Indeed, this is the single most important feature that allows directional detectors to gain sensitivity to the WIMP signal compared to backgrounds.  Since spin independent WIMP scattering is azimuthally symmetric about the
direction of the incoming WIMP the dominant WIMP signal comes from the
a comparison of forward-backward scattering events.  This is seen in the results in
Tables~\ref{tab:Nevents-100} and~\ref{tab:Nevents-1000}.  A three
dimensional detector, even with perfect angular resolution, but without forward-backward discrimination requires
at least an order of magnitude more events than a three dimensional detector with such discrimination and even
many more than a poorly aligned two dimensional detector to distinguish a WIMP signal from terrestrial backgrounds.  This is because without forward-backward discrimination the detector relies upon the
difference between head-on and glancing (wide angle) collisions as well as
high angular resolution to distinguish a WIMP signal from the background.

\begin{figure}
  \includegraphics[angle=-90,width=3in]{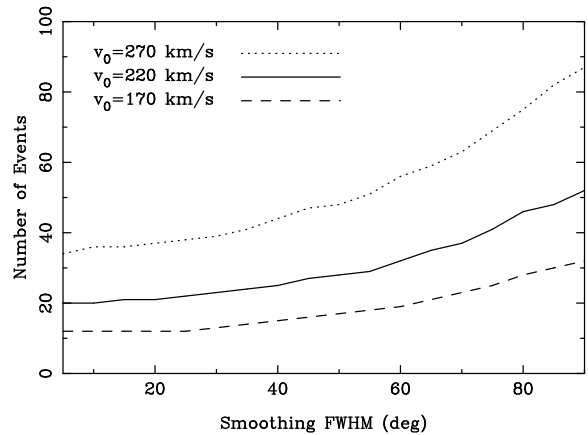}
  \caption{The number of events required as a function of the full-width
    half maximum (FWHM) of the smoothing
    kernel~(\ref{eqn:smoothing-kernel})
    for the isothermal models and $m_\chi=100\GeV$.}
  \label{fig:Nevents-smooth}
\end{figure}

We next explore how the sensitivity of a three-dimensional detector
depends upon its angular resolution. In figure~\ref{fig:Nevents-smooth} we
display the number of events required as a function of the angle of the
full-width half maximum (FWHM) of the detection cone for the events. Note that
as the angular resolution degrades, the number of events required for a
three dimensional detector quickly approaches that of a two-dimensional
detector, as expected. In order to be significantly more efficient, the
angular resolution of such a detector must be better than about 60
degrees (fwhm).

We finally focus on two-dimensional detectors, in part because these are likely to be the most practical in the near future, and because less attention in the past has been paid to them than hypothetical three dimensional detectors.

\begin{figure}
  \includegraphics[angle=-90,width=3in]{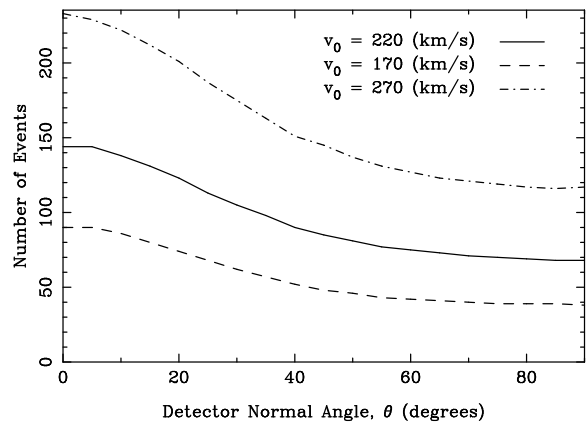}
  \caption{The number of events required as a function of the angle between the
    detector normal and the direction of the WIMP wind (Sun's motion),
    $\theta$ for the isothermal models and $m_\chi=100\GeV$.}
  \label{fig:Nevents-2D}
\end{figure}

It is clear that the efficacy  of a planar detector will depend upon the orientation of its plane with respect to the direction of the WIMP wind.
Specifically a two-dimensional detector fixed to the Earth will be oriented so that its
normal vector makes an angle $\theta$ with the Earth's rotation axis.  The
choice of $\theta$ determines how much time the detector will spend at
various angles $\zeta$ relative to the Earth's direction of motion.  An
orientation of $\zeta = 0^\circ$ is clearly the worst since the detector
plane is then perpendicular to the WIMP wind.   The
number of events required is a function of the angle $\theta$ chosen for
the detector as shown in figure~\ref{fig:Nevents-2D}.    

Tables~\ref{tab:Nevents-100} and
\ref{tab:Nevents-1000} give the minimum and maximum number of events
required for optimal versus worst-case orientation of the detector.  Note that the shape of the function in
figure~\ref{fig:Nevents-2D} depends on the orientation of the Earth's axis,
$\epsilon = 42^\circ$ relative to the motion of the Sun through an isotropic halo.  For halos in which a WIMP stream arose which was not due to the motion of the Sun, the shape of the curve would change.  Low values of $\epsilon$  in this case would produce a more
steeply descending function, and high values would produce an ascending
function.  

Note that the results thus far assume that the detector axis is fixed to the plane of the Earth.  In this case the time averaged number of events required is on average 3 times less for a two-dimensional detector oriented in the best possible axis (ie.  at or near $\theta = 90^\circ$
for isothermal halo distribution).  Such a detector
however requires three to four times the number of events of a full
three-dimensional detector.    Whether or not achieving this additional factor of 3 in sensitivity for a full three-dimensional detector is worth the technological challenge is not clear.  However, as seen in the tables, this factor of three can be reduced  by using  a detector which is not fixed relative to the plane of the Earth, but which can rotate over the course of each day with respect to the earth to maintain an optimal orientation with respect to an expected WIMP wind.  For an isothermal halo, roughly only 2 times as many events are required for such a rotating detector compared to a detector with full three-dimensional tracking capability.  The technical difficult of producing the former may be less demanding than that required to produce the latter. 

Our results can be summarized as follows:  For directional WIMP
detectors, forward-backward discrimination is far more valuable than
three-dimensional resolution of the track, at least for the isothermal
model considered here.  Furthermore, a two-dimensional detector can be
oriented into the predominant WIMP wind so that the number of events required to distinguish a WIMP halo from a terrestrial background is comparable to that required for even a full three dimensional detector.   While there may be other reasons (including background rejection) for considering fully three dimensional directional detection methods, our results thus suggest that concentrating simply on forward-background sensitivity is the most important new direction that should be pursued in directional WIMP detection, and that planar detectors can, in this case, provide nearly optimal detectional sensitivity  (We note that as this work was being prepared for submission a related study appeared which complements this work, and is consistent with our results~\cite{directional-statistics-2d}.).

%\begin{figure*}
%  \resizebox{5in}{!}{\rotatebox{-90}{\includegraphics{dama_phase.ps}}}
%  \caption{Modulation phase for non-symmetric halo distributions.  For the
%    standard model the phase is $152.5\days$.}
%  \label{fig:phase}
%\end{figure*}

\bibliographystyle{apsrev}
\bibliography{angular}

\end{document}